\documentclass[aps,prc,twocolumn,floatfix,showpacs,preprintnumbers,amsmath,amssymb,nofootinbib,groupedaddress]{revtex4-1}
\usepackage{color}
\usepackage{graphicx}
\usepackage{dcolumn}    
\usepackage{multirow}
\usepackage{bm}         
\usepackage{sidecap}
\usepackage{hyperref}   
\usepackage{xcolor}
\usepackage{amsmath}
\usepackage{hyperref} 
\usepackage{soul}
\definecolor{dark-red}{rgb}{0.,0.,0}
\definecolor{dark-blue}{rgb}{0.,0.,1}
\definecolor{medium-blue}{rgb}{0,0,1}
\hypersetup{
    colorlinks, linkcolor={dark-red},
    citecolor={dark-blue}, urlcolor={medium-blue}
}

\begin{document}
\title{Charge Transport in \emph{Bifidobacterium animalis subsp. lactis BB-12} under various atmospheres}

\author{K. Bozkurt\textsuperscript{1}}
\email{kutsalb@yildiz.edu.tr}
\author{C. Denkta{\c{s}}\textsuperscript{1}}
\author{O. {\"{O}}zdemir\textsuperscript{1}}
\author{A. Alt{\i}ndal\textsuperscript{1}}
\author{Z. Avdan\textsuperscript{2}}
\author{H. S. Bozkurt\textsuperscript{3}}

\affiliation{\textsuperscript{1}Yildiz Technical University, Davutpasa Campus, Science and Arts Faculty, Physics Department, 34220 Esenler Istanbul, Turkey\\
             \textsuperscript{2}Kocaeli University, Umuttepe Campus, Science and Arts Faculty, Physics Department, 41380 Izmit Kocaeli, Turkey \\				
               \textsuperscript{3}Maltepe University, Medical Faculty Internal Medicine, Clinic of Gastroenterology, 34843 Maltepe Istanbul, Turkey
               }

\date{\today} 

\begin{abstract}
The influence of relative humidity (RH) on quasistatic current-voltage ${(I-V)}$ characteristics of \emph{Bifidobacterium animalis subsp. lactis BB-12} thin layers was studied for the first time. The value of electrical conductivity in 75$ \%$ RH was found to be in the order of 10$^{-7}$ (ohm cm)$^{-1}$ \textcolor{red} {,} which was 10$^{6}$ orders of magnitude higher than that observed in dry atmosphere. It was concluded that RH played a key role in hysteresis behaviour of the measured ${(I-V)}$ characteristics. FTIR measurements showed that under water moisture environment the associated bonds between amine and carboxyl group were greatly strengthened that was the source of free charge carries after ionization. The surface charge of \emph{Bifidobacterium animalis subsp. lactis BB-12} was found to be negative by zeta potential measurements, claiming that electrons were the charge carriers.
\end{abstract}

\maketitle
\section{Introduction}
\subsection{\emph{BB-12} and Gut Microbiota}
The gut microbiota contains a diverse community of commensal, symbiotic and pathogenic microorganisms \cite{1,2}. The gut microbiota has anti-inflammatory, antioxidant, antioncogenic effects and it contributes to the immunological, hormonal and metabolic homeostasis of the host \cite{3,4}. The genus \emph{Bifidobacterium} belongs to the phylum \emph{Actinobacteria} and it comprises Gram-positive, non-motile, often branched anaerobic bacteria \cite{5}. The \emph{Bifidobacteria} are one of the major species in the human colon microbiota and have been frequently used as probiotic \cite{6}. \emph{Bifidobacterium animalis subsp. lactis BB-12} is a catalase-negative, rod-shaped bacterium which was first isolated in 1983 (Figure-1). At the time of isolation, \emph{Bifidobacterium animalis subsp.} was considered as one of the species of \emph{Bifidobacterium bifidum} \cite{7}.

In 2010, the complete genome sequence of BB-12 was mapped \cite{7}. The BB-12 genome consists of a single circular chromosome of 1,942,198 base pairs with 1642 predicted protein-encoding genes, 4 rRNA operons, and 52 tRNA genes \cite{8}. A physical mapping of the BB-12 chromosome revealed that the genome sequence was correctly assembled (Figure-2).
\begin{figure}
  \centering
  \includegraphics[width=0.4 \textwidth]{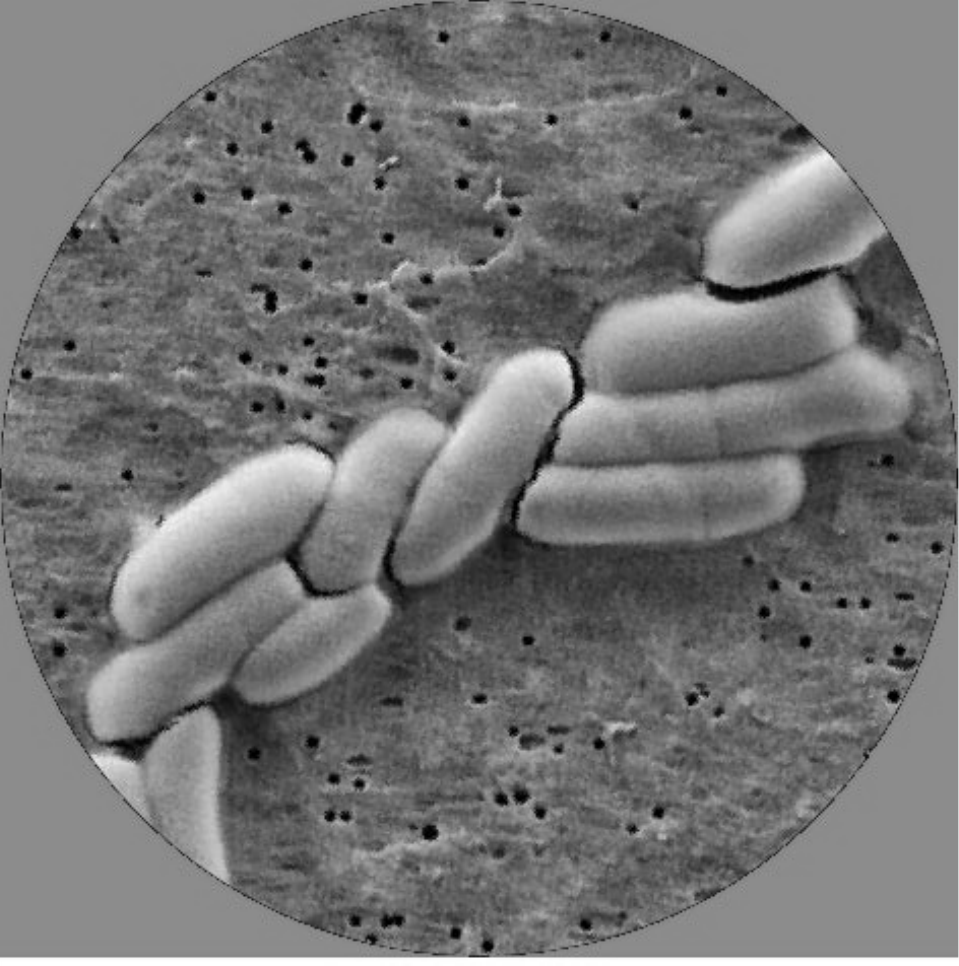}
  \caption{ \emph{Bifidobacterium animalis subsp. lactis BB-12 (Chr Hansen) }  strain. }
\end{figure}
\begin{figure*}
  \begin{center}
\includegraphics[width=0.6\linewidth,clip=true]{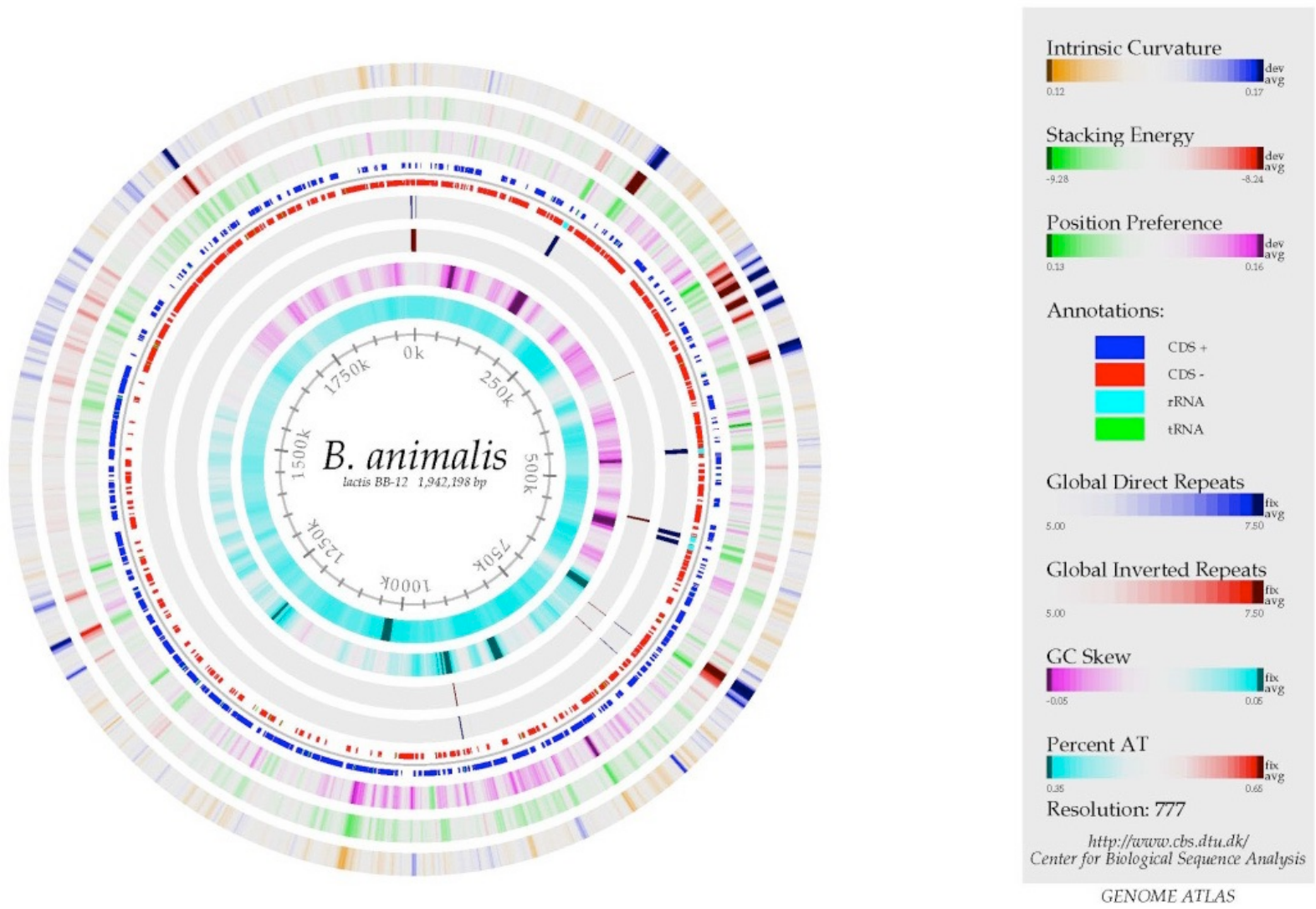}
  \end{center}
  \caption{The BB-12 genome atlas. The physical mapping of the BB-12 chromosome revealed that the genome sequence was correctly assembled. From Garrigues et al. 2013 \cite{8} with permissions from Elsevier, Copyright 2013.}
  \label{601-}
\end{figure*}
\begin{figure}
  \centering
  \includegraphics[width=0.5 \textwidth]{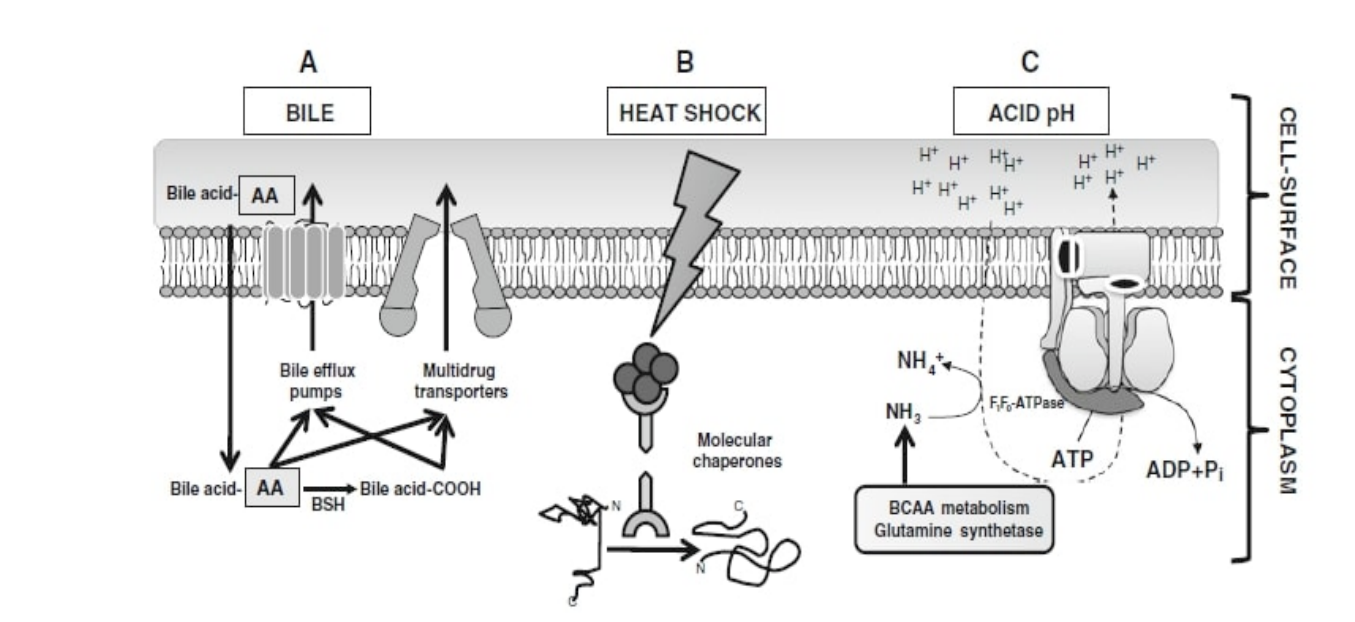}
  \caption{ Main molecular mechanisms involved in the response of BB-12 to different stresses. A Bile is detoxified from the cytoplasm by the activity of bile efflux pumps and/or multidrug transporters. Conjugated bile acids are deconjugated by the bile salt hydrolase, although the relationship of this enzyme with the resistance to bile is unclear. B Both bile and heat shock induce protein aggregation and misfolding, which is counteracted by the action of chaperones and proteases. C The F1F0-ATPase is used by bifidobacteria for counteracting the cytoplasm acidification that occurs in acidic environments. In addition, production of branched-chain amino acids is coupled with glutamine deamination, rendering ammonia that acts as a cytoplasmic buffer. From Lorena et al \cite{9} with permissions from Springer-Verlag , Copyright 2011.}
\end{figure}	
BB-12 is technologically well suited, expressing fermentation activity, high aerotolerance, good stability and a high acid and bile tolerance \cite{8}. Because of high redox potential in the colon flora ecosystem, BB-12 is highly resistant against acidic pH, digestive enzymes and  toxic effect of  bile acids \cite{8}. The BB-12 cell envelope is an electrical and physical barrier that can be overcome by pathways that consist of redox proteins and structural proteins. Some of the main molecular mechanisms and electron transport systems are presented in Figure 3.

 \subsection{\emph{BB-12} and Bacterial Cellular Electron Transport Systems}	
	
Bacterial cellular electron transport systems (CET) are defined as microbial bioelectrochemical processes in which electrons are transferred from the cytosol to the membrane of the cell \cite{10}. Shi, L. et al showed molecular mechanisms that underlie the ability of bacterial to exchange electrons, such as c-type cytochromes and microbial nanoparticles between bacterial membrane and gut enterocytes \cite{10}. Samuel, H. Light et al described food borne gut pathogen L. \emph{monocytogenes} cellular electron transfer systems \cite{11}. They showed that NADH dehydrogenase enzymatic pathway is the responsible mechanism for  CET from aerobic respiration by channelling electrons to a discrete membrane-localized quinone pool. Although Lorena et al showed the strength of BB-12 under lower pH, bile acid and digestive enzymes conditions, CET features of BB-12 were still not completely identified. Here we first described the probiotic bacteria \emph{Bifidobacterium animalis subsp. lactis BB-12} strain features of CET under  various atmospheres. This work was organized as follows. Materials, Zeta-potential studies, ATR-FTIR analysis, SEM and SEM images electrical characterization of BB-12 were presented in Sec.II. Charge transport system of \emph{Bifidobacterium animalis subsp. lactis BB-12} was discussed in Sec.III. Conclusions and outlooks were given in Sec. IV.	

\section{Materials and Methods}	
\subsection{Material}	
Mean gut bifidobacterial count of 5 billions ${(5 \times 10^9)}$ cfu of \emph{Bifidobacterium animalis subsp. lactis BB-12}  (Chr Hansen) were used in this study. This \emph{Bifidobacterium} was dissolved in 5 $ml$ of distilled water and the sample was prepared. Spin coating technique was employed to deposit the films of the \emph{Bifidobacterium animalis subsp. lactis BB-12} onto the interdigital micro electrode (IDE) arrays. For the spin coating processes, the coating solutions were prepared by dissolving the BB-12 in double distilled water at concentrations of $\sim$1 M. 200 $\mu L$ of such solution was added with a glass pipette onto the IDE structure held onto spinner (Speciality Coatings Systems Inc., Model P6700 Series). The substrate was spun at 2500 rpm for 60 s and then the film was transferred to the test chamber.
\subsection{Zeta Potential}		
The Zeta potential (ZP) of BB-12 was measured in water using a Zetasizer Nano ZS (Malvern Instruments Ltd) at room temperature. The voltage applied to the driving electrodes of the capillary electrophoresis cell was $149 V$. The zeta potential of bacteria was calculated based on the Smoluchowski equation and the values of three different samples were averaged, 
\begin{equation}
\xi=\frac{dU}{dp}\frac{\mu}{(\varepsilon\times\varepsilon_0)}\times \kappa
\end{equation}
where $\xi$ is the zeta potential, $dU/dP$ represents the slope of the streaming potential versus pressure, $\mu$ is the viscosity, $\varepsilon$ is the relative permittivity of the electrolyte solution, $\varepsilon_0$ is the electric permittivity of vacuum and $\kappa$ is the electrolyte conductivity.
\subsection{FTIR Spectral Method }	
The change in structure of BB-12 was recorded by an Agilent Technologies Cary 630 FTIR apparatus at room temperature. Spectra of all samples were recorded between 400 and 4000 cm$^{-1}$ at a resolution of 4 cm$^{-1}$ with 100 scans.	
\subsection{Electrical Characterization of \emph{Bifidobacterium animalis subsp. lactis BB-12}}		
Photolitographically patterned interdigital arrays of tin (Sn) electrodes on plexiglass substrate were used for the electrical characterization of \emph{Bifidobacterium animalis subsp. lactis BB-12}. The interdigital electrodes consisted of 20 finger pairs of electrodes with a width of 100 $\mu$m and a space of 100  $\mu$m between adjacent electrodes. Then, the BB-12 film surface exposed to the various levels of RH between 0 $\%$ and 100 $\%$ for 10 min. After each exposure, curret-voltage characteristics of the film was measured using an electrometer (Keithley model 617) which was connected to personal computer by an IEEE 488 data acquisition interface. During the deposition of the BB-12 film, the temperature of the interdigitated substrate was kept at 300 K. A home made stainless steel test chamber with a capacity of $1\times10^{-5}$ liters was used in these experiments. Dry nitrogen was used as carrier gas and the desired relative humidity inside the test chamber was obtained by bubbling dry nitrogen gas through double distilled water. Well defined levels of relative humidity were obtained by mixing dry nitrogen gas and water vapor using a computer driven mass flow controllers (Alicat Scientific, Inc.). Direct current (dc) conductivities $(\sigma)$ of the BB-12 film was calculated from the measured current-voltage $(I-V)$ characteristics by using following equation;
\begin{equation}
\sigma=\frac{I}{V}\frac{d}{(2m-1)L h}
\end{equation}
where $I$ is the measured current, $V$ is the bias voltage, $d$ is the electrode spacing, $m$ is the number of electrode finger pairs, $L$ is the overlap length of the electrode fingers and $h$ is the thickness of the electrodes. 

\section{Results and Discussion}	
\subsection{Zeta-potential studies}	
The zeta potential of BB-12 was measured using a Zeta master (Malvern Instruments, Malver, UK) at room temperature. The zeta potential for BB-12 was found to be negative $(-7.85 mV)$, which indicated that the cell surface was predominantly covered with anionic compounds, such as strong acids, the phosphate based (lipo-) teichoic acids, weak acids, the carboxylate containing acidic polysaccharides and proteins \cite{12}. This negative value is in accordance with the previous studies performed considering that (i) CIDCA 5310, 537 and NCC 189 grown in bile-containing medium \cite{13} and (ii) free L. rhamnosus GG cells as a function of pH \cite{14}. Moreover, similar range of zeta potentials have also been reported for microencapsulation in Alginate and Chitosan Microgels to Enhance Viability of \emph{Bifidobacterium} \cite{15}. Then, the zeta potential of BB-12 was increased from $(-7.85 mV)$ to $(-10.4 mV)$ as a function of time, which can be an indication that the acidic character of the surface was reduced \cite{12}.
\begin{figure*}
  \begin{center}
\includegraphics[width=0.85\linewidth,clip=true]{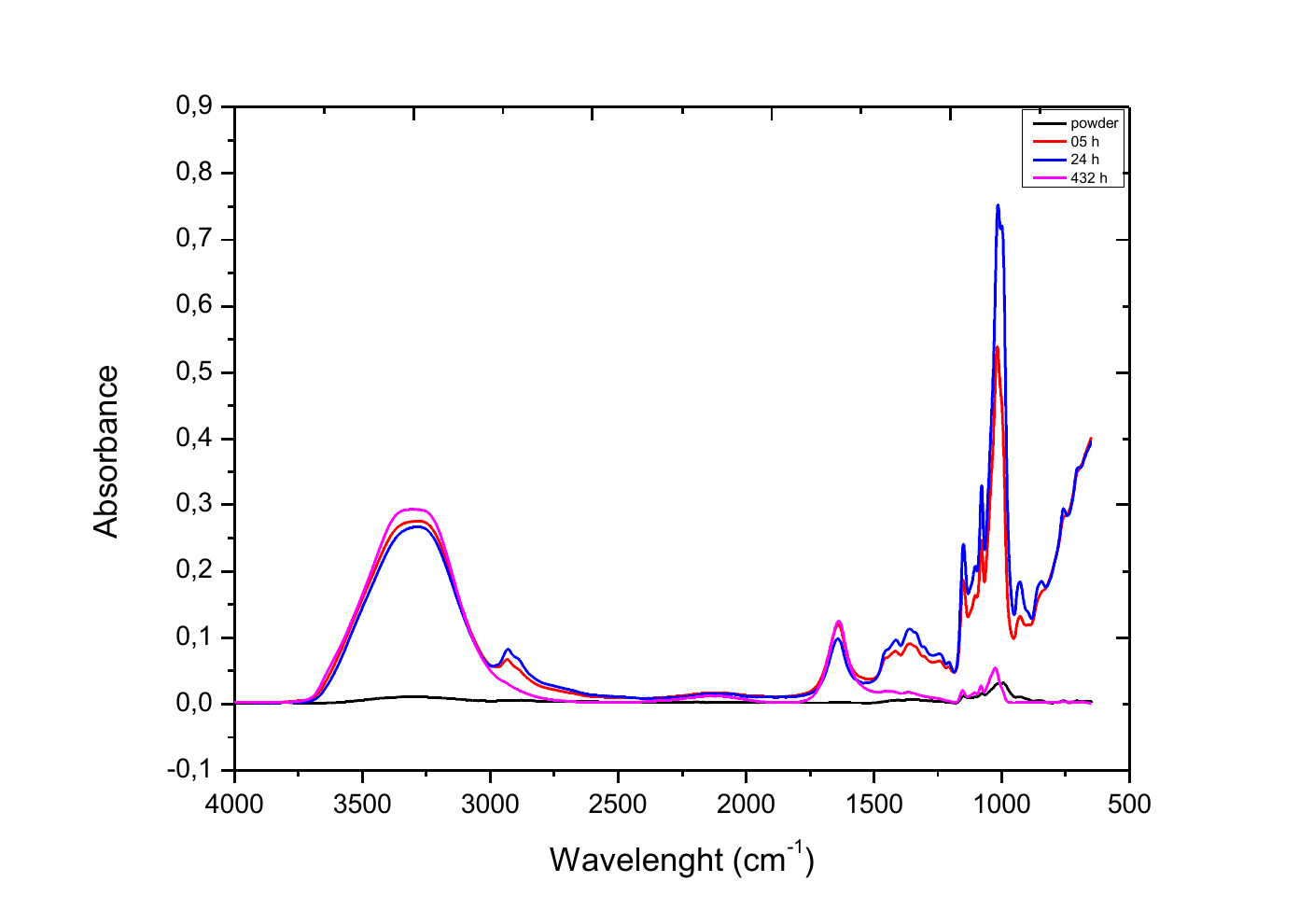}
  \end{center}
  \caption{ FTIR spectra of \emph{Bifidobacterium animalis subsp. lactis BB-12} with the water exposure of time.}
  \label{601-}
\end{figure*}
\subsection{ATR-FTIR Analysis}	
The surface structure of the cells of BB-12 were examined by ATR-FTIR spectroscopy. The changes in the structure of all the samples were recorded with the time of water exposure and spectrum is shown in Fig. 4. New bonds and increase in the bond intensity were found in the BB-12 spectra after treatment of distilled water.  Powder BB-12, a small broad band at 3301.81 $cm^{-1}$ was observed, this band intensity increased after treatment distilled water, which corresponds to the hydroxyl stretching vibration of the polysaccharide \cite{16,17}. The vibrational small peak at 2910.32 $cm^{-1}$ is related to the alkyl-hydrocarbons groups, the so called fatty acid region \cite{18,19}. After contacting with the pure water, an increase in the intensity of this peak was observed \cite{18,19}. As shown in Fig.4, the very small peak at 1636.62 $cm^{-1}$ is related to the carbonyl stretching of secondary amides (Amide $I$) \cite{18,20}. But, it was observed that after distilled water treatment the increase for the amide $I$ absorption band.
The very small peaks at 1407 and 1354 $cm^{-1}$ are assigned to the bending of $-CH_{3}$ and $-CH_{2}$ groups of bacterium \cite{20}. The borad and intensive peaks were observed between 900-1200 $cm^{-1}$ refers to polysaccharides and carbohydrates groups and the peak 995 $cm^{-1}$ was due to the valence $C-O-C$ group vibrations in the cyclic structures \cite{17,18,19}. But, after 432 h of distilled water treatment of carbohydrate compounds on the surface decreased. The band located the at 852 $cm^{-1}$ is attributed to indicate that configurations exist in the polysaccharides \cite{16}. The change of spectral peaks is in agreement with the zeta-potential results of a BB-12 that has passed into the endogenous phase.
\subsection{SEM analysis}
\begin{figure}
  \centering
  \includegraphics[width=0.52 \textwidth]{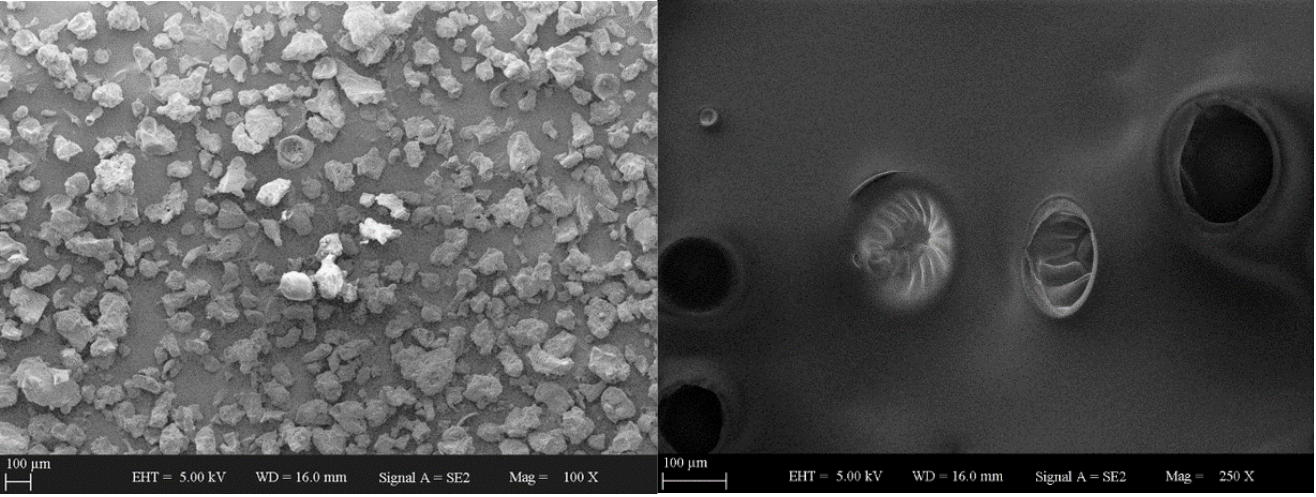}
  \caption{ SEM surface micrograph of \emph{Bifidobacterium animalis subsp. lactis BB-12} with untreated (left) and treated (right).}
\end{figure}
The surface morphologies of powder BB-12 before treatment with distilled water were characterized using a scanning electron microscope (SEM) and the results are presented in Figure 5. It is seen in Fig. 5 that the bacterial integrity is maintained and non-spherical structure before purified water. 
\begin{figure*}
  \begin{center}
\includegraphics[width=1.05\linewidth,clip=true]{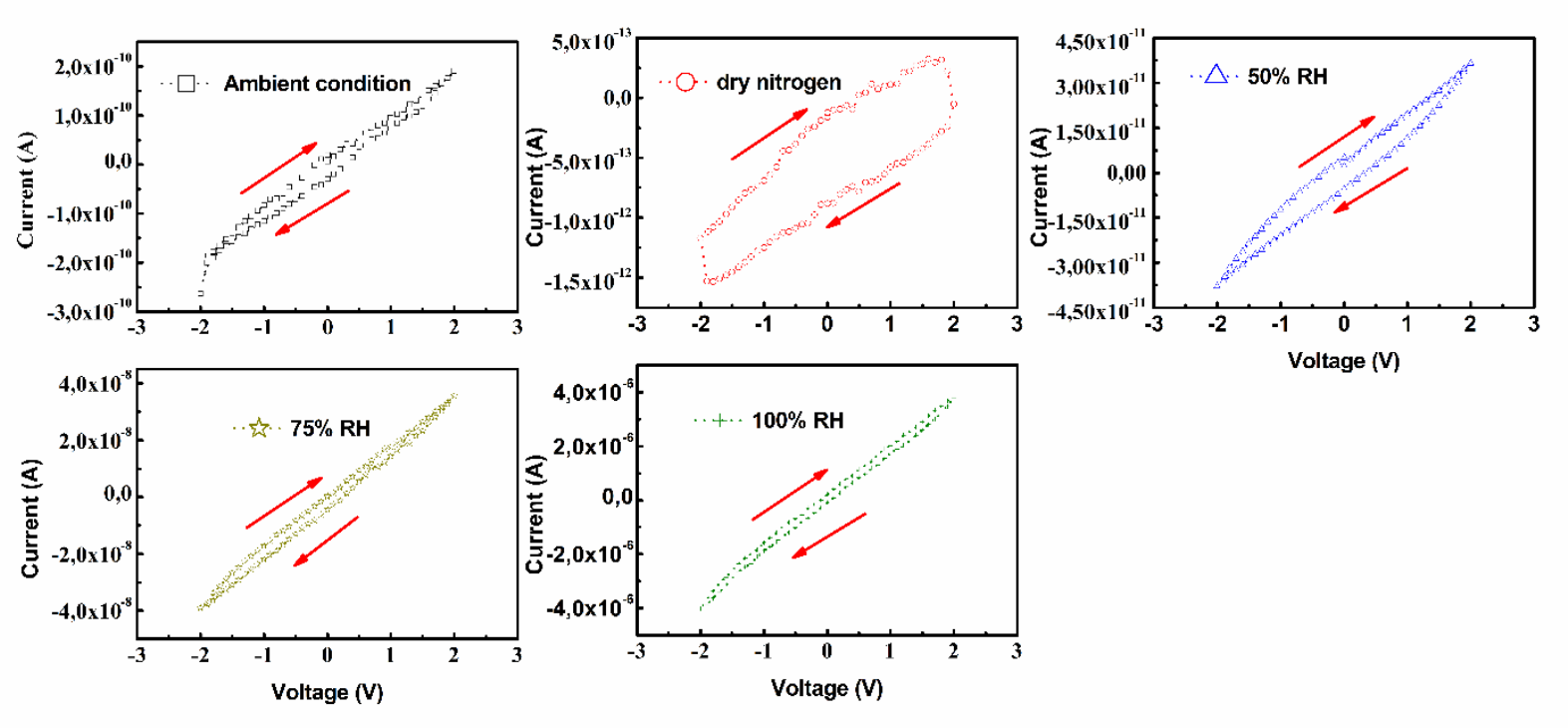}
  \end{center}
  \caption{$I-V$ characteristics of the film of \emph{Bifidobacterium animalis subsp. lactis BB-12} at various RH levels.}
  \label{601-}
\end{figure*}	
\subsection{Charge transport in \emph{Bifidobacterium animalis subsp. lactis BB-12}}	

Charge transport in BB-12 film under direct current condition was investigated by current-voltage $(I-V)$ measurements. The measurements were carried out at room temperature under various RH conditions.  Fig. 6 shows the $I-V$ characteristics of the sample at indicated RH levels. In this measurement the voltage was incremented in steps of $50 mV$ from $-1$ to $+1$ $V$ and back again. The measured $I-V$ curves exhibited considerable hysteresis between increasing and decreasing voltage sweeps. It is important to point out that the area within the loop decreased with an increase in RH level. The interpretation of appearance of the hysteresis behavior in dry nitrogen atmosphere and low RH level requires to take into account many elementary physical processes, such as density of mobile charges, and delay time between two successive measurements separated by a potential step. A considerable increase in film electrical conductivity with the increase in RH level, compared to dry nitrogen atmosphere, was also clearly. Observed the effect of the film conductivity was reversible for all RH levels investigated, which indicated that when the film surface was purged with carrier gas (dry nitrogen) the conductivity of the film returned to the initial value. The relative humidity level was controlled with a commercially available humidity meter and it was found to be less than 2 $\%$. Increase in dc conductivity of the film revealed that the interaction between the water molecules and the BB-12 film was based on charge transfer. When the water molecules interacted with the surface of the BB-12 film, BB-12 exhibited n-type character, the zeta potential became more negative due to the formation of functional groups such as $COO$ and $NH_{2}$. 
	
\section{Conclusions}

Charge transport behavior of the BB-12 film and the effect of the RH level on it were investigated by means of $I-V$ measurements. Within the relative humidity, electrical conductivity of the BB-12 increased more than six decades while under $N_{2}$ environment conductivity returned to the initial current value. This behaviour in conductivity modulation was reversible at least in the three cycles. These experimental findings showed that there was no structural transformation at different relative humidity levels. On the other side, increase in the conductivity was observed with the increase in the population of charge carries, supplied by the interaction of BB-12 with the water moisture, monitored through amine and carboxyl group by FTIR and Zeta potential measurements. Overall, the obtained results n this study indicated that \emph{Bifidobacterium animalis subsp. lactis BB-12} has a great potential for humidity detection at room temperature. This type of environment control of the conductivity for the conductivity of BB-12 will pave the way for many working areas such as biomaterials and biosensors.


%

\end{document}